\documentclass[3p,times,twocolumn]{elsarticle}

\usepackage{ecrc}

\usepackage{graphicx}
\usepackage{amsmath,amssymb}

\volume{00}

\firstpage{1}

\journalname{Nuclear Physics B Proceedings Supplement}

\runauth{}

\jid{nppp}

\jnltitlelogo{Nuclear Physics B Proceedings Supplement}

\usepackage{amssymb}

\usepackage[figuresright]{rotating}

\begin{document}

\begin{frontmatter}




\title{Jet Hadronization via Recombination of Parton Showers in Vacuum and in Medium}


\author{Rainer J.\ Fries}
\author{Kyongchol Han}
\author{Che Ming Ko}

\address{Cyclotron Institute and Department of Physics and Astronomy, Texas
  A\&M University, College Station TX 77845, USA}

\begin{abstract}
We have studied the hadronization of jet parton showers based on the quark 
recombination model. This is achieved by letting gluons at the end of the 
perturbative shower evolution undergo a non-perturbative splitting into 
quark and antiquark pairs, then applying a Monte-Carlo version of 
instantaneous quark recombination, and finally subjecting remnant quarks 
(those which have not found a recombination partner) to Lund string 
fragmentation. When applied to parton showers from the PYTHIA Monte Carlo 
generator, the final hadron spectra from our calculation compare quite well 
to PYTHIA jets that have been hadronized with the default Lund string 
fragmentation. Modeling the quark gluon plasma produced in heavy ion 
collisions by a blast wave model, we have further studied medium effects 
on the hadronization of jet shower partons by also including their 
recombination with the thermal partons from the quark gluon plasma. We find 
that the latter leads to a significant enhancement of intermediate transverse 
momentum pions and protons at both RHIC and LHC. Our results thus suggest that 
medium modification of jet fragmentation provides a plausible explanation for 
the enhanced production of intermediate transverse momentum hadrons observed 
in experiments.
\end{abstract}


\end{frontmatter}

Jets in quantum chromodynamics (QCD) have a long history as tools to test
QCD itself, electroweak theory, and physics beyond the standard model. 
Recent developments both in theory and experiment have also made jets into 
promising probes in heavy ion physics. Jets embedded in quark gluon 
plasma (QGP) created in high energy nuclear collisions suffer from 
jet quenching. Details of the jet-medium interaction encode important 
aspects of QGP at various scales. To make better connections
between theory and experiment several groups around the world are currently
developing event-by-event jet shower Monte Carlo (MC) simulations. Those 
shower MCs are reasonably controlled as long as the shower is
in the perturbative domain, typically determined by the virtuality of partons
$Q$, or the temperature $T$ of the surrounding medium. As $Q$ reaches a 
lower cutoff, usually around 1 GeV, and the temperature reaches 
the critical temperature $T_c$ perturbative methods have to be discarded.
This leaves the hadronization of partons in jet showers unaccounted for.

Non-perturbative models like the Lund string fragmentation model used in
PYTHIA \cite{Sjostrand:2006za} or the cluster hadronization model used in 
HERWIG \cite{Corcella:2002jc} have been used  to describe the transition
from partons to hadrons in jets in the vacuum, e.g.\ in $e^++e^-$ or $p+p$ 
collisions. On the other hand, there is strong evidence in 
A+A collisions, that recombination of partons from jets with partons 
in the surrounding medium can create hadrons. The 
instantaneous quark recombination model
\cite{Greco:2003xt,Fries:2003vb,Greco:2003mm,Fries:2003kq,Fries:2004ej,Fries:2008hs} 
has been applied both to the recombination between thermal partons, and for 
recombination between thermal partons and leading jet partons.
The instantaneous recombination model has been successfully deployed 
for a variety of observables at intermediate momenta ($\sim$ 2-6 GeV/$c$)
in heavy ion collisions. Clearly, the possibility that quarks in jet showers
pick up partons from the thermal medium to form hadrons, as they pass through 
the $T=T_c$ hypersurface in the collision, needs to be accounted for. Such
a mechanism involves the exchange of momentum, energy and flavor quantum 
numbers of the jet shower with the medium, thus influencing a number of
jet and high-$p_T$ hadron observables. It needs to be studied carefully
theoretically, and in light of available experimental jet reconstruction 
techniques.

\begin{figure}[htb]
\begin{center}
\includegraphics[width=0.9\columnwidth]{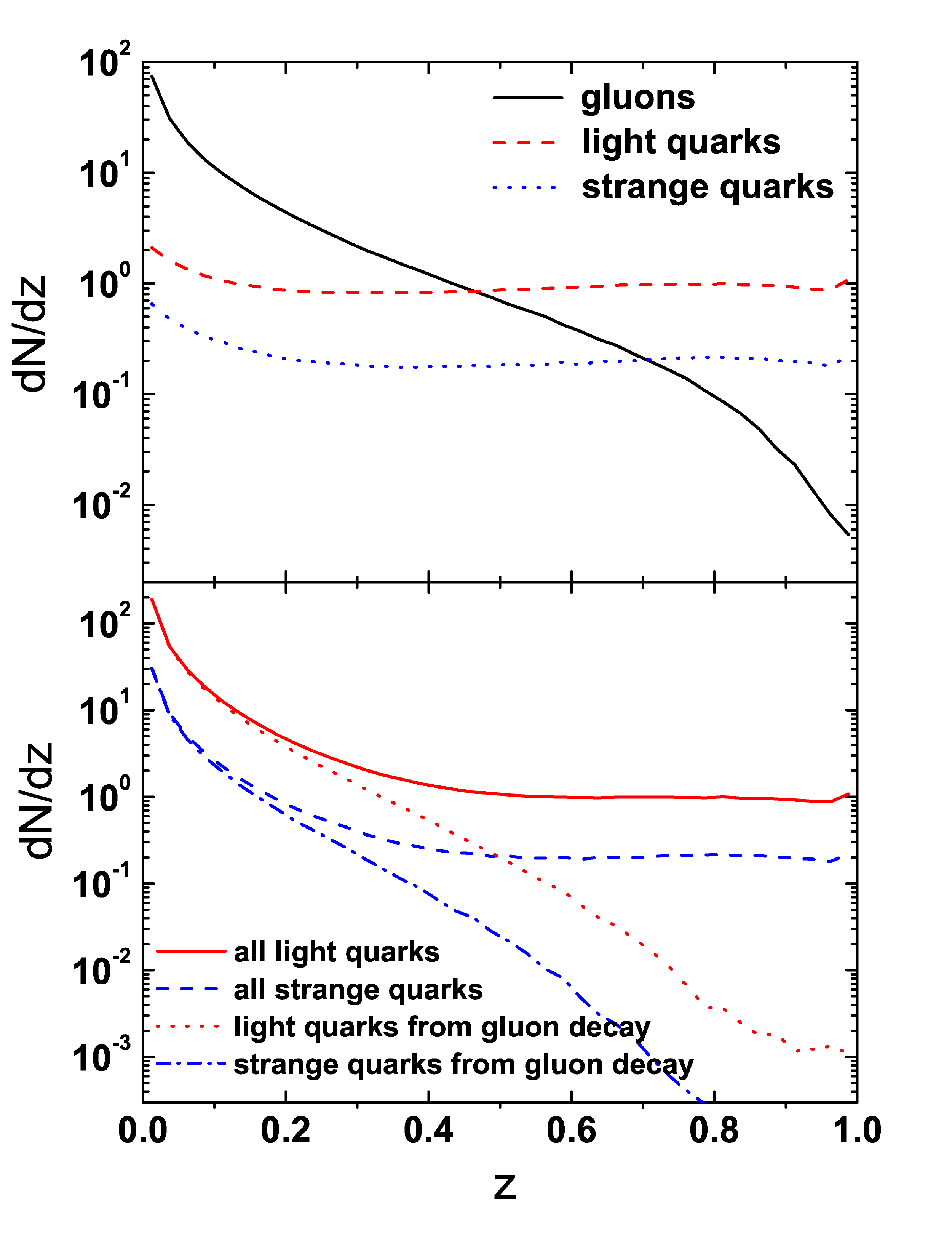}
\caption{Distribution $dN/dz$ of partons in 100 GeV PYTHIA quark showers
 before (top panel) and after (bottom panel) decay of gluons.}
\label{fig:1}
\end{center}
\end{figure}

Here we report on our effort within the JET collaboration to develop a
hadronization module for jet shower MCs that incorporates quark recombination
effects for jets in a medium but also reproduces hadronization of jets
in the vacuum. This will be accomplished by a hybrid approach using 
recombination and string fragmentation. The idea that quarks in a jet
shower could hadronize by recombination has earlier been discussed in 
\cite{Hwa:2003ic}. Let us start by considering
a perturbative jet shower in the vacuum. All partons have been evolved to some
small virtuality $Q_0$ of order 1 GeV. For test purposes we will use
$e^++e^-$ showers produced by PYTHIA 6 \cite{Sjostrand:2006za}. The first step toward
hadronization is a forced splitting of gluons into quark-antiquark pairs.
This is done by simply decaying gluons isotropically in their rest frame
into light quarks $u\bar u$, $d\bar d$ or $s\bar s$ using their remnant 
virtualities. The resulting distribution of partons in the jet shower in
terms of their momentum fraction $z$ of the total momentum before and 
after the decay is shown in Fig.\ \ref{fig:1}.

The instantaneous recombination model projects quark states onto hadron states
enforcing only momentum conservation. This can be conveniently expressed 
in terms of the Wigner functions of the quark-antiquark (3-quark) Wigner 
function and a meson (baryon) Wigner function \cite{Fries:2008hs}. In the
case of mesons
\begin{multline}
\label{Eq1}
\frac{dN_M}{d^3 {\mathbf p}_M} = g_M\int d^3{\bf r}_1 d^3{\bf k}_1 d^3{\bf r}_2
d^3{\bf k}_2 f_{q}({\bf r}_1, {\bf k}_1)
f_{\bar{q}}({\bf r}_2, {\bf k}_2) \\ \times W_{M}(\Delta {\bf r}, \Delta {\bf
  k})\delta^{(3)}({\bf p}_{M}-{\bf k}_1 -{\bf k}_2) \, ,
\end{multline}
where $W_M$ is the meson Wigner function, $\Delta \mathbf{r}$ and $\Delta
\mathbf{k}$
are the relative position and momentum of the partons, $g_M$ is a statistical 
factor, and the quark-antiquark Wigner function has been factorized into single 
quark Wigner functions $f$. The latter is an approximation neglecting 
correlations in the parton system. A similar formula can be devised for
baryons \cite{Fries:2008hs,Han:new}.

\begin{figure}[htb]
\begin{center}
\includegraphics[width=0.9\columnwidth]{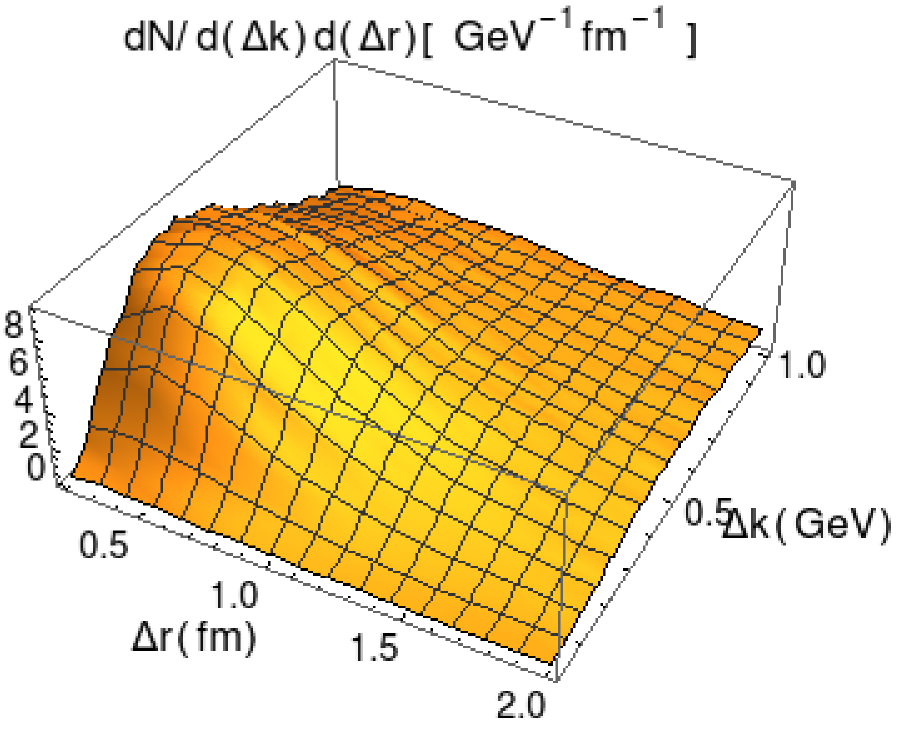}
\caption{The distribution of quark-antiquark pairs as a function of their
distances $\Delta r$ and $\Delta k$ in coordinate and momentum space, resp.
Distances are measured in their common rest frame. 100 GeV PYTHIA showers
have been used.
  }
\label{fig:2}
\end{center}
\end{figure}

It is clear that the rate of hadronization through recombination will 
depend on the density of partons in phase space, more precisely on the
average distance in phase space relative to the width
of the hadron Wigner function in phase space. Fig.\ \ref{fig:2} shows
the distribution of quark-antiquark pairs in 100 GeV PYTHIA showers
(after gluon decays) as a function of distances $\Delta r$ and 
$\Delta k$ in position and momentum space. Those distances are measured 
for each pair in its common rest frame at the time the later parton is 
created.
We can see that the distribution peaks at around $\Delta r \approx 0.5$ fm
and $\Delta k \approx 0.4$ GeV. Thus many quarks in the shower could in fact
find another parton in rather close proximity. On the other hand, long
tails exist in the distribution which indicate the existence of 
``lonely quarks'' which are unlikely to find a recombination partner.

We can now proceed to calculate the recombination probability for all
pairs (and triplets) in a shower, and use MC techniques to pick recombining 
mesons (and baryons). To first approximation one can use spherical wells as
hadron Wigner functions as in \cite{Greco:2003xt}. A more realistic approach 
postulates an harmonic oscillator approximation for quark bound states.
The Wigner function of the $n^{\mathrm{th}}$ excited state ($n=0$ corresponds to the
ground state) of a 1-D harmonic
oscillator of frequency $\omega$ is well known to be \cite{Groenewold:1946kp}
\begin{eqnarray}
W_n (u)= 2(-1)^n L_{n} \left(\frac{4u}{\hbar\omega}\right)e^{-2u/\hbar\omega},
\end{eqnarray}
where $u=\frac{\hbar\omega}{2}\left(\frac{x^2}{\sigma^2}+\sigma^2 k^2\right)$,
$\sigma\equiv\left(\frac{\hbar}{m\omega}\right)^{1/2}$, and the $L_n$ are 
Laguerre polynomials. The Wigner function of a single quark in this bound
state has to be smeared by at least the minimal quantum mechanical uncertainty
to yield a positive Wigner function. With Gaussian wave packets of width
$\delta = \sigma/\sqrt{2}$ (a choice made for simplicity \cite{Han:new}) we 
obtain for the $n^{\mathrm{th}}$ state a probability
\begin{equation}
\psi_n = \left(\frac{u}{\hbar \omega}\right)^n \frac{1}{n!} e^{-u/\hbar \omega}
\end{equation}
The construction of states for the 3-D case are then straight forward
\cite{Han:new}.
We fix the width parameters $\sigma$ for the stable hadrons $\pi$, $K$ and $N$
by using measured charge radii. We allow recombination into excited states
which are currently given by the excited harmonic oscillator spectrum,
with a subsequent decay into stable hadrons.

\begin{figure}[htb]
\begin{center}
\includegraphics[width=1.0\columnwidth]{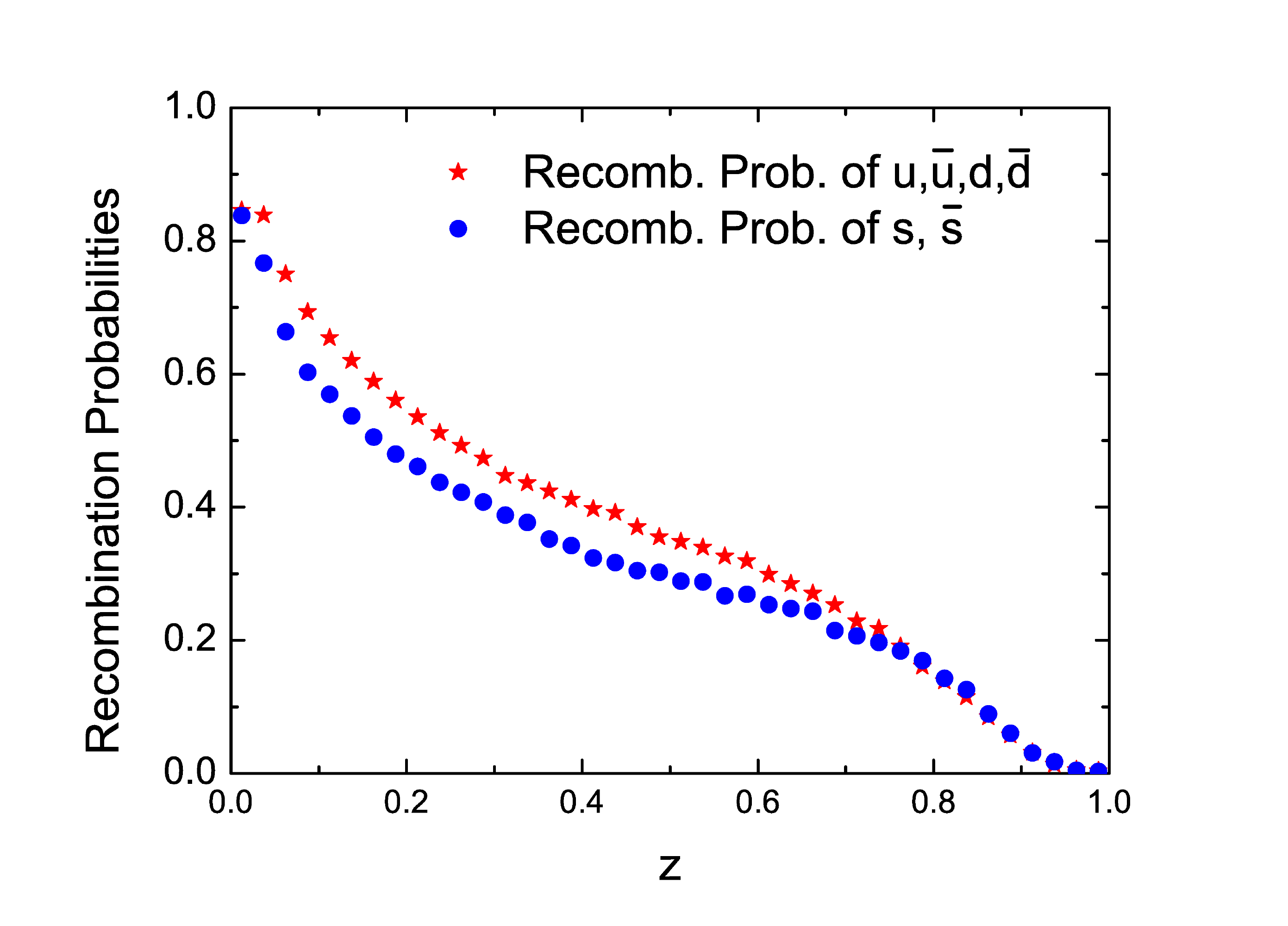}
\caption{The fraction of quarks undergoing quark recombination as a function of
momentum fraction $z$ for 100 GeV jet showers.}
\label{fig:2a}
\end{center}
\end{figure}

\begin{figure}[htb]
\begin{center}
\includegraphics[width=1.0\columnwidth]{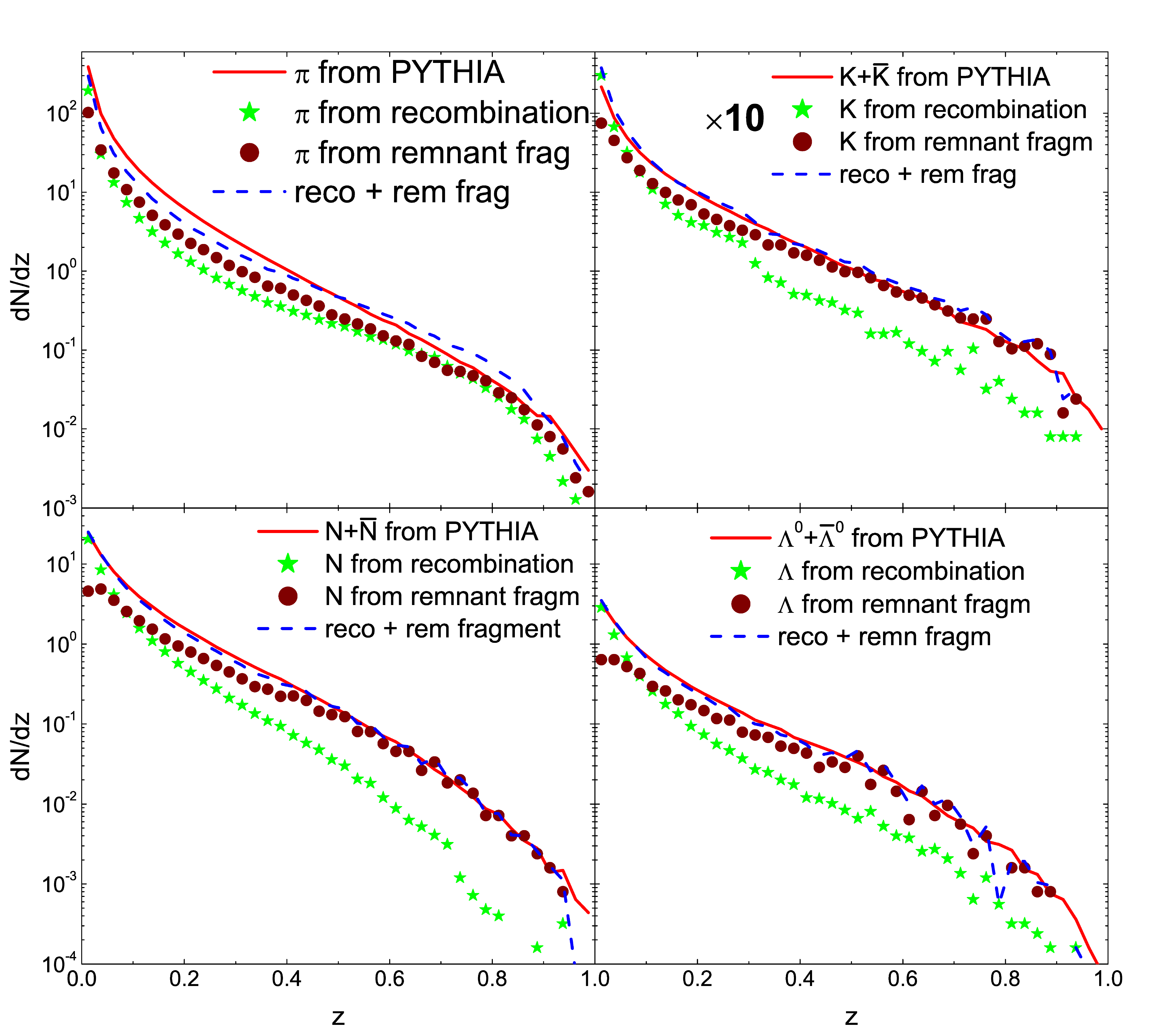}
\caption{The distribution $dN/dz$ of hadrons in 100 GeV PYTHIA quark showers
 from pure string fragmentation (PYTHIA), and the model discussed here
 (reco+rem frag). The contributions from recombination and string
 fragmentation in our model are also shown separately. Clockwise from 
 the upper left panel: pions, kaons, $
 \Lambda$s and nucleons.}
\label{fig:3}
\end{center}
\end{figure}

When we recombine mesons and baryons in jet showers as described we find
that a large fraction of small-$z$ partons indeed finds a recombination
partner, while at larger $z$ the success of recombination drops rapidly,
see Fig.\ \ref{fig:2a}. These
remnant partons have propagated far away in phase space from other quarks.
In reality, instead of facing the QCD vacuum they would rather be forced 
to connect to another color charge via a QCD string. Hence we reintroduce 
strings between remnant partons. We use PYTHIA 6 string fragmentation 
routines to calculate the hadrons from this remnant string fragmentation. 
This completes the brief overview of our hadronization module. All partons 
in the system have been converted into hadrons.

\begin{figure}[htb]
\begin{center}
\includegraphics[width=1.0\columnwidth]{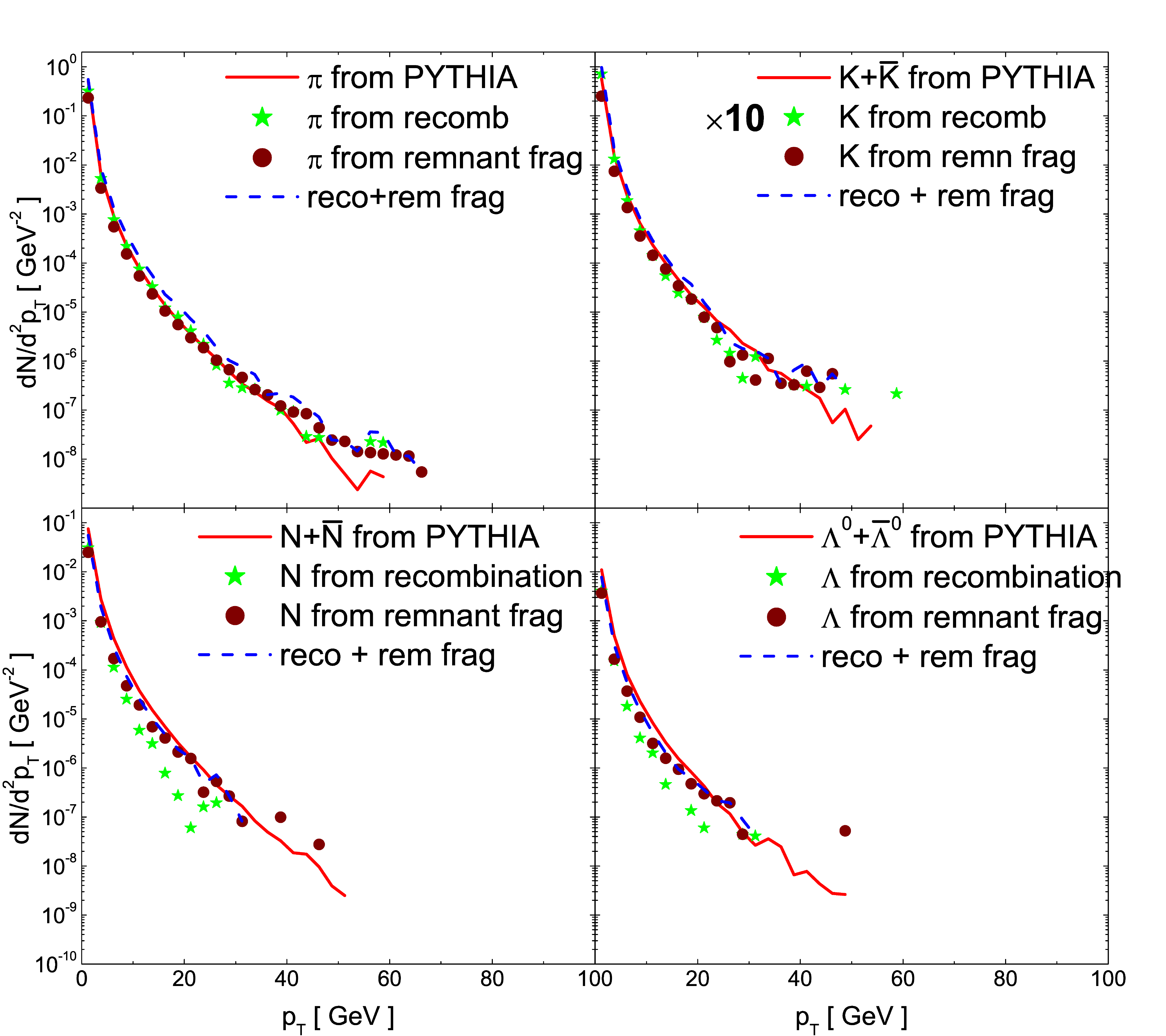}
\caption{The same as Fig.\ \ref{fig:3} but the distribution $dN/d^2 p_T$ 
 of momentum transverse to the jet axis is shown.}
\label{fig:4}
\end{center}
\end{figure}

Fig.\ \ref{fig:3} shows hadron spectra from our recombination-fragmentation
hybrid model applied to 100 GeV PYTHIA parton showers, compared to the same
parton showers hadronized directly with string fragmentation in PYTHIA. 
We see that the distribution of pions, kaons, nucleons and $\Lambda$s as
a function of momentum fraction $z$ generally coincide rather well with
the results of pure string fragmentation. Fig.\ \ref{fig:4} shows the same
for the distribution of momenta transverse to the jet axis. We conclude that
our hybrid hadronization model which replaces string fragmentation 
by quark recombination for partons close together in phase space is well
suited to describe fragmentation of jets in the vacuum. In the near future
we plan to test the module directly with data in $e^++e^-$ and $p+p$
collisions.

Why would one want to replace a successful hadronization model like
string fragmentation by quark recombination? The answer is that
implementing medium effects is rather straight forward in recombination 
models. In addition to considering the recombination probabilities of
quarks and antiquarks in a jet shower one also samples the thermal partons
from a blastwave or from the $T=T_c$ hypersurface of a fluid dynamic 
simulation, and allows pairs and triplets between thermal partons and shower
partons as well. For individual jets the details of the location
relative to the hypersurface is very important and the relative importance
of shower-thermal and shower-shower recombination depends on this geometry.
Note that in a typical situation (a jet originating a few fm inside the 
$T=T_c$ hypersurface) the bulk of the jet shower partons are created 
inside the QGP. Thus they have to propagate through the medium to the 
hadronization hypersurface, or are absorbed by the medium, processes which
will have to be carefully simulated in shower MCs. However, the 
leading part of the jet typically emerges from the hypersurface, i.e.\ 
these partons are born outside of the QGP.

Using a blast wave model we find that overall shower-thermal 
recombination (for baryons the mixed terms, shower-shower-thermal and
shower-thermal-thermal, exist) in most situations becomes more important 
than shower-shower recombination and leads to a noticeable enhancement of
pion and proton production at intermediate hadron $p_T$. 
The increase in shower-thermal recombination gives rise to phenomena like
the proton/pion enhancement observed in experiment.
In the near future the in-medium realization of our hadronization module will
be combined with realistic in-medium jet shower MCs to study these effects in
detail.

In summary, we have developed a hybrid hadronization model which employs
quark recombination of shower partons and uses string fragmentation
to hadronize remnant quarks. We find that our model can reproduce 
the spectra of pure string fragmentation in the vacuum rather well. 
The introduction of 
medium partons is rather straight forward and gives rise to hadrons from 
shower-thermal recombination. In collaboration with groups developing
in-medium jet shower MCs simulations we will apply our hadronization model
to the phenomenology of jets and high-$p_T$ hadrons in heavy ion collisions
in the near future.

RJF would like to thank the organizers of the Hard Probes 2015 conference
for choosing this work for presentation.
This work was supported by the U.S. National Science Foundation by CAREER
Award PHY-0847538 and grants PHY-1516590 and PHY-1068572, the US Department of Energy under Contract DE-FG02-10ER41682 within the frame of the JET Collaboration, and the Welch Foundation under Grant No.\ A-1358.









\end{document}